\documentclass[aps,prl,twocolumn,superscriptaddress,showpacs]{revtex4}
\usepackage{amsmath, amsfonts, amssymb, graphicx,epstopdf}
\newcommand{\bra}[1]{\langle #1 |}
\newcommand{\ket}[1]{| #1 \rangle}
\newcommand{\braket}[2]{{\langle  #1 | #2 \rangle}}
\newcommand{\braketop}[3]{\langle #1 | #2 | #3 \rangle}
\newcommand{\ave}[1]{\left\langle #1 \right\rangle}
\newcommand{\dg}[0]{\dagger}

\newcommand{\vk}[0]{{\bf k}}

\newcommand{\vq}[0]{{\bf q}}
\newcommand{\vQ}[0]{{\bf Q}}

\newcommand{\vR}[0]{{\bf R}}

\newcommand{\hexaneg}[0]{{\! \! \! \! \! \!}}
\newcommand{\tr}[0]{{\rm tr}}
\newcommand{\twodots}[0]{{..}}

\renewcommand{\epsilon}{\varepsilon}

\newcommand{\footnoteremember}[2]{
\footnote{#2}
\newcounter{#1}
\setcounter{#1}{\value{footnote}}
\hspace{-10pt}
}
\newcommand{\footnoterecall}[1]{
\footnotemark[\value{#1}]
\hspace{-7pt}
}

\begin{document}
\title{Quasiparticle Theory of Resonant Inelastic X-ray Scattering in High-T$_c$ cuprates}

\author{David Benjamin}
\affiliation{Physics Department, Harvard University, Cambridge, Massachusetts, USA}

\author{Israel Klich}
\affiliation{Department of Physics, University of Virginia, Charlottesville, VA, USA}

\author{Eugene Demler}
\affiliation{Physics Department, Harvard University, Cambridge, Massachusetts, USA}

\date{\today}

\begin{abstract}
We develop a formalism for calculating resonant inelastic x-ray scattering (RIXS) spectra in systems of itinerant electrons with arbitrary band structures, accounting for the effect of the positively-charged core hole exactly.   We apply this formalism to the cuprate superconductors and obtain quantitative agreement with experimental data over a wide range of dopings.  We reproduce the dispersing peaks and non-trivial polarization dependence found in several experiments.  Thus we explain by band structure alone features previously attributed to collective magnetic modes.
\end{abstract}

%PACS
%Scattering, x-ray in condensed matter: 78.70.Ck
%Hole-doped cuprates: 74.72.Gh
\pacs{78.70.Ck, 74.72.Gh}
%\keywords{}

\maketitle

Resonant inelastic x-ray scattering (RIXS) is unique among energy-resolved probes of electronic excitations in its ability to measure momenta over most of the Brillouin zone.  It couples to a wide variety of excitations~\cite{Hill2008, Braicovich2010,LeTacon2011,Lee2013, Dean2013,Ulrich2009, Schlappa2012} and, due to large flux and bulk sensitivity, does not suffer from limitations of sample size and surface quality.  In its intermediate state a core electron is raised to an excited state and this electron's dynamics give a ubiquitous lowest-order contribution to RIXS spectra.   Although RIXS experiments have been peformed over a wide range of doping, most theoretical work has focused on Mott insulating phases~\cite{Brink2007, Vernay2008, Chen2010} in parent materials and cluster models~\cite{Vernay2008,Jia2013}.  Consequently, data from materials with itinerant electrons have been interpreted in terms of models of insulators.

\begin{figure}
\includegraphics[width=\linewidth]{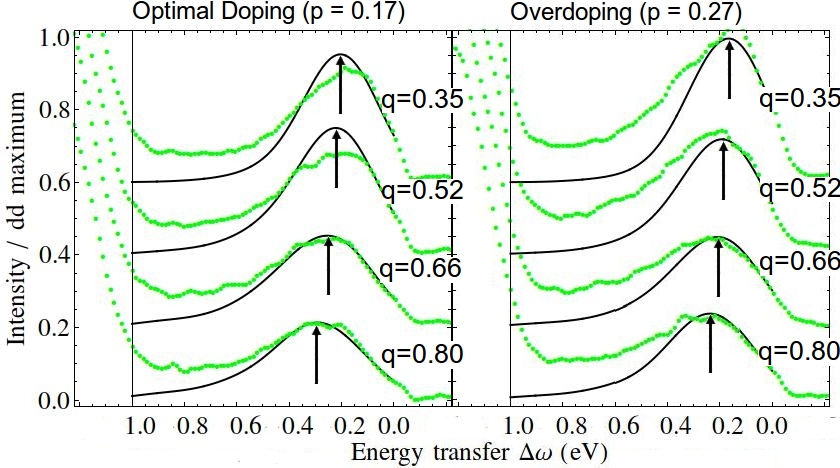} 
\caption{Calculated (black curve) and experimental (Ref.~\cite{LeTacon2013}, green dots) intensity vs. energy transfer for spin-flip RIXS of optimally-doped and overdoped Tl-2201 for antinodal momenta $\vQ = q (\pi/a, 0)$ exhibiting identical dispersing peaks.  We subtracted a non-resonant elastic peak from the raw experimental data.  We did not remove the contribution of $dd$ excitations, which accounts for the discrepancy at large $\Delta \omega$.} 
\label{Lineshapes}
\end{figure}

In this paper we calculate RIXS spectra using a model of non-interacting quasiparticles but including an interaction with a positively-charged core hole via an exact determinantal method.  We derive formulas for both direct and indirect RIXS, which differ by which band the core electron is raised to.   For direct RIXS, we account for spin-orbit splitting of the core level, which opens a spin-flip (SF) channel in addition to the non-spin-flip (NSF) channel~\cite{Ament2009, Ament2011, Haverkort2010}.  We apply our formalism to cuprates over a range of doping and achieve quantitative agreement with experimental data (Fig.~\ref{Lineshapes}).  In particular, peaks in the calculated and measured lineshapes disperse identically.  As in experiments we also find that NSF lineshapes are broader and higher in energy than SF lineshapes.  These features were previously attributed to magnetic effects, but we find that band structure alone produces dispersing lineshapes, while the core hole combines with Pauli blocking to separate SF and NSF lineshapes.

\begin{figure}
\includegraphics[width=\linewidth]{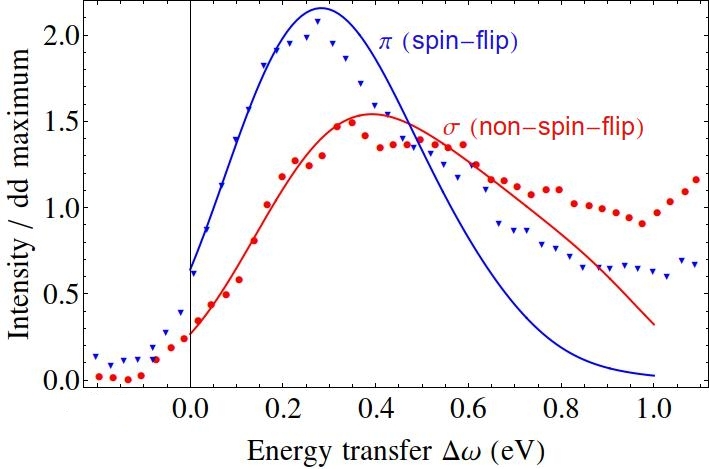} 
\caption{Spin-flip ($\pi$-polarized) and non-spin-flip ($\sigma$-polarized) antinodal $\vQ = 0.80 (\pi/a, 0)$ lineshapes of optimally-doped ($p=0.15$) Bi-2212 for core hole potential $U_c = 1.0$ eV.  Blue: spin-flip channel/$\pi$-polarization; red: non-spin-flip channel/$\sigma$-polarization.  Solid lines: calculated results; triangles, circles: $\pi$- and $\sigma$- polarized data from Ref.~\cite{Dean2013a}.}
\label{Polarization}
\end{figure}

{\it Theoretical Formalism}.--  Incident photons $\vq$, $\omega$ scatter into outgoing state $\vq+\vQ, \omega-\Delta \omega$ with intensity~\cite{Kotani2001, Ament2011, Haverkort2010} $I \propto \sum_f \left| A_f \right|^2 \delta(E_f - E_i - \Delta \omega)$, where
\begin{equation}
\label{amplitude}
A_f = \sum_m e^{i \vQ \cdot \vR_m} \chi_{\rho \sigma} \braketop{f}{d_{m \rho} (H_m + \omega - E_i + i \Gamma)^{-1} d^\dg_{m \sigma}}{i}.
\end{equation}
Here $\ket{i,f}$ and $E_{i,f}$ are the initial and final electron states and energies, $d^\dg_{m\sigma}$ creates a valence electron of spin $\sigma$ at site $m$, and $\chi_{\rho \sigma}$ is a polarization-dependent $2 \times 2$ spin matrix that comes from the product of two dipole matrix elements: $\braketop{3d_\sigma}{T\sigma}{2p}$ and $\braketop{2p}{T_\rho}{3d_\rho}$ for absorption and emission, where $T$ is a dipole transition operator\footnoteremember{supp}{We explicitly derive the effects of spin-orbit coupling in the Supplemental Material.  See also Refs.~\cite{Ament2011} and \cite{Haverkort2010}}\hspace{-4pt}.  In direct RIXS the strong ($\sim$ 20 eV) spin-orbit coupling of the $2p$ core level implies that the energy eigenstates $\ket{2p_{J,m}}$ are not eigenstates of spin $S_z$.  Thus off-diagonal elements $\rho \ne \sigma$ are permitted.  In indirect RIXS the core hole is in a $1s$ state and its spin is conserved, hence $\chi_{\sigma \rho}$ is diagonal. One can isolate either the diagonal (NSF) or off-diagonal (SF) component of $\chi$ by varying only the incident polarization\footnoterecall{supp}.  The immobile intermediate state core hole does not appear explicitly in Eq.~(\ref{amplitude}) but affects the valence system indirectly.  It forces absorption and emission to occur on the same site $m$ and contributes a width $\Gamma$ and a potential $V_m$ acting on valence electrons, $H_m = H+V_m$, to the intermediate state.    
%In a basis where $\hat{z}$ lies in the cuprate CuO$_2$ plane, $\chi_{\up \up} = \chi_{\dn \dn}$ and $\chi_{\up \dn} = \chi_{\dn \up}$.

Due to the core hole the eigenstates of $H_m$ have no simple relation to those of $H$ and it is convenient to work in the time domain.  This puts the intensity in the form~\cite{Nozieres1974}
\begin{align}
\label{TimeDomain}
I \propto &  \int_{-\infty}^\infty \hexaneg ds \int_0^{\infty} \hexaneg dt  \int_0^{\infty} \hexaneg d\tau \, e^{i\omega (t-\tau)-is \Delta \omega-\Gamma (t+\tau)}  \nonumber \\
& \times \sum_{mn} e^{i \vQ \cdot (\vR_m-\vR_n)} \chi_{\rho \sigma} \chi_{\mu \nu} S_{\rho \sigma \mu \nu}^{mn}, \\
S_{\rho \sigma \mu \nu}^{mn} =&
 \left\langle e^{i H \tau} d_{n \rho} e^{-i H_n \tau} d^\dg_{n \sigma} e^{i H s} \ldots \right. \nonumber \\
\label{S}
& \quad \quad \left. d_{m \mu} e^{i H_m t} d^\dg_{m \nu} e^{-i H(t+s)} \right\rangle.
\end{align}
One obtains Eq.~(\ref{TimeDomain}) via the identities $1/z=\int_0^\infty e^{-zt} dt$, $\delta(z) = \int e^{i s z} ds$, replacing eigenvalues by operators, and recognizing resolutions of unity, e.g. $\sum_f \ket{f}\bra{f} \ldots \ket{i} \delta(E_f-E_i-\Delta \omega)$ $\rightarrow$ $\int ds \, e^{-i \Delta \omega s} e^{i H s} \ldots e^{-i H_i}\ket{i}$.  However, Eq.~(\ref{S}) is best understood as a history of absorption and emission events separated by time evolution operators, that is, as the time-dependent amplitude to scatter a photon.  The intensity is obtained from the square of this amplitude, hence the pair of creation and annihilation operators is followed by its Hermitian conjugate.  

In the following analysis we treat the valence band as a system of non-interacting quasiparticles.  This approximation is valid when the quasiparticle lifetime is long compared to the core hole lifetime $1/\Gamma$, in which case an electron is unlikely to be scattered by other electrons in the brief time between absorption and emission.  In the cuprates, for example, typical values are $\Gamma = 300 -500$ meV, while quasiparticle widths are smaller than this even quite far from the the Fermi surface.  (We stress that negligible scattering on short time scales is logically distinct from a Fermi liquid ground state~\cite{Deng2013}; the former, for example, implies nothing about DC transport).  We further assume that the high-energy band in indirect RIXS is highly dispersive and non-interacting, so that the core electron excited into this band is a ``spectator" to the interaction of the valence band with the core hole.

In indirect RIXS of the cuprates and other transition metal oxides, where a $1s$ core electron is raised to a $4p$ band, the $4p$ photoelectron's dynamics reduce to a Green function $d_m e^{i H_m t} d^\dg_m \rightarrow  \braketop{0}{d_m e^{i H_{4p} t} d^\dg_m}{0} e^{i H_{3d,m} t} \equiv G^{mm}_p(-t) e^{i H_{3d,m} t}$, where $H_{4p}$ and $H_{3d,m}$ are the Hamiltonians of the $4p$ band and the valence $3d$ band with a core hole at $m$.  Because the $1s$ core hole has no spin-orbit coupling, $\chi_{\rho \sigma} \propto \delta_{\rho \sigma}$ and the RIXS process is effectively spinless.  Hence
\begin{align}
&S^{mn} = G^{nn}_p(\tau) G^{mm}_p(-t) \times \nonumber \\
\label{IndirectS}
& \quad \ave{e^{i H \tau}  e^{-i H_n \tau}  e^{i H s}  e^{i H_m t}  e^{-i H(t+s)}}.
\end{align}
We simplify the many-body average in Eq.~(\ref{IndirectS}) in terms of the single particle matrices $h_{(m,n)}$, where lowercase letters denote the matrix elements of a quadratic operator: $H = d^\dg_i h_{ij} d_j$~\cite{Abanin2005, Benjamin2013}.  With $N \equiv \left(1+ e^{\beta h } \right)^{-1}$ we obtain
\begin{align}
&S^{mn}= G^{nn}_p(\tau) G^{mm}_p(-t) \det \left[ \left( 1-N\right) \right. \nonumber \\
\label{SmnFormula}
&\left. +  e^{i h \tau}  e^{-i h_n \tau}  e^{i h s}  e^{i h_m t}  e^{-i h (t+s) } N\right].
\end{align}
To compute $S^{mn}_{\rho \sigma \mu \nu}$ for direct RIXS we extend a method applied to tunneling in quantum wires~\cite{Abanin2005} and resonant elastic x-ray scattering~\cite{Benjamin2013}, which involved matrix elements like those in Eq.~(\ref{S}) but with one $d$ and one $d^\dg$.  We present the  straightforward but lengthy derivation in the appendix.  The result is
 \begin{align}
\label{super}
 &S^{mn}_{\rho \sigma \mu \nu} = \det(F) \left[
 \braketop{n\rho}{(1-N)F^{-1} e^{-i h_n \tau}}{n\sigma} \right. \nonumber \\
 & \times \braketop{m\mu}{e^{-i h s} e^{i h_n \tau} (1-N) F^{-1} U_{mn}}{m\nu}   \nonumber \\
 &+ \braketop{n\rho}{(1-N)F^{-1} U_{mn}}{m\nu} \nonumber \\
 &\times \left. \braketop{m\mu}{e^{i h_m t} U_0 N F^{-1} e^{-i h_n \tau}}{n\sigma} \right].
 \end{align}
 where 
 $U_{mn} = e^{-i h_n \tau}e^{i h s} e^{i h_m t}$, $U_0 = e^{i(\tau-t-s)h}$, and $F = 1-N + U_{mn} U_0 N$.  For a full band $S$ vanishes, as it should, and for an empty band it reduces to $\braketop{n}{e^{-ih_n \tau}}{n} \braketop{m}{e^{i h_m t}}{m}$, a general term in the the expression $\left| \sum_m \int  G^{mm}_{3d}(t) dt \right|^2$.  That is, the amplitude of RIXS in an empty band is the coherent sum of electron propagators that start and end at the same core hole site.  Eq.~(\ref{super}) pertains to a full spin-orbital basis, but for a spin-independent Hamiltonian easily factorizes.  If $H$ contains a singlet pairing term $d^\dg_{m,\uparrow} B_{mn} d^\dg_{m,\downarrow}$ it  can be put into non-anomalous form suitable for matrix manipulations via a transformation $d^\dg_{m \uparrow} \leftrightarrow d_{m \uparrow}$.  This handles all spin density waves and pairing terms that occur in the cuprates.  More complex spin density waves and triplet pairing require a more sophisticated formalism~\footnote{I. Klich, unpublished}.

{\it Results}.--  We now apply this formalism to study cuprate superconductors, comparing our results to experiments on Tl$_2$Ba$_2$CuO$_{6+\delta}$ (Tl-2201) and Bi$_2$Sr$_2$CuO$_{6+x}$ (Bi-2212).  An outstanding puzzle is the existence of peaks in direct RIXS not seen in neutron scattering~\cite{Wakimoto2007} or indirect RIXS~\cite{Hill2008}.  We take $H=\sum_{\vk, \sigma} \epsilon_\vk d^\dg_{\vk,\sigma} d_{\vk,\sigma}$, where $\epsilon_\vk = -2 t_1 (\cos (k_x) + \cos (k_y)) - 4 t_2 \cos (k_x) \cos (k_y) - 2 t_3 (\cos (2 k_x) + \cos (2 k_y) ) - 4 t_4 (\cos (2 k_x) \cos (k_y) + \cos (k_x) \cos (2 k_y) )$, using canonical tight-binding band structures fit to ARPES data: $(t_1, t_2, t_3, t_4) = (126, -36, 15, 1.5)$ meV for Bi-2212~\cite{Markiewicz2005} and $(t_1, t_2, t_3, t_4) = (181, -75, -4, 10)$ meV for Tl-2201~\cite{Peets2007}.  We assume an attractive contact potential $V_m = -U_c \sum_\sigma d^\dg_{m \sigma} d_{m \sigma}$ for the core hole. We fix $\omega$ at the absorption maximum as in experiments.  Fig.~\ref{Lineshapes} shows SF intensity versus $\Delta \omega$ for antinodal momenta $\vQ \parallel (\pi,0)$ in optimally-doped and overdoped Tl-2201 ($p=0.17$ and $p=0.27$) with $U_c = 1.0$ eV along with data from Ref.~\cite{LeTacon2013}.  We subtracted a Gaussian elastic peak at $\Delta \omega = 0$ from all experimental data and convolved calculated lineshapes with Gaussians of width equal to the instrumental resolutions of the corresponding experiments.  We choose $U_c=1.0$ eV to obtain the best fit to NSF lineshapes; SF RIXS is nearly independent of $U_c$.   The most striking feature is an intensity peak that disperses to higher energy with increasing momentum, reaching a maximum of $250 - 300$ meV, as seen in experiments~\cite{LeTacon2011,Dean2012, Dean2013, LeTacon2013,Dean2013a}.  One possible interpretation is that these peaks are due to inelastic scattering of a collective mode.  However, we see that band structure alone can produce them \footnote{Zeyher and Greco make a similar claim about Raman spectra in Ref.~\cite{Zeyher2013} and argue that RIXS behaves similarly.}.   Quantitatively, the calculated and experimental lineshapes agree very well, with the location of peaks and their low-energy side in nearly perfect agreement.    There is a systematic discrepancy at large values of $\Delta \omega$ due to the tail of orbital $dd$ excitations~\cite{Zaanen1990}.  It is reassuring that this discrepancy is nearly independent of momentum, as local excitations ought to be.  The calculated nodal and antinodal lineshapes also agree very well with experimental data from optimally-doped Bi-2212 \footnote{H. Ronnow, unpublished}.   

In Fig.~\ref{Polarization} we show that the agreement between theory and experiment extends to NSF scattering.  This is important because the SF and NSF channels correspond to spin and charge degrees of freedom and a difference in their lineshapes is seen as compelling proof of magnetic physics.  Indeed, the matrix elements in Eq.~(\ref{SimpleDirect}), below, are manifestly spin-independent, so that SF and NSF lineshapes should be identical in the absence of interactions.   However, the core hole potential dramatically separates SF and NSF lineshapes.  As $U_c$ increases, the NSF peak moves to higher energies and broadens while the SF peak remains relatively sharp, exactly as seen in experiments. For $U_c \sim 1.0$ eV the agreement is very good up to energies at which the $dd$ tail becomes significant.

The core hole separates SF and NSF lineshapes as follows:  Its attractive potential tends to keep the photoexcited electron of spin $\sigma$ bound near $\vR_m$, leading to elastic scattering.  Pauli blocking prevents other electrons of spin $\sigma$ from hopping onto $\vR_m$ and filling the core hole, thereby robbing spectral weight from inelastic scattering.  With sufficient energy the photoexcited electron may be dislodged, allowing inelastic scattering.  Hence NSF scattering with small $\Delta \omega$ is suppressed relative to scattering with large $\Delta \omega$.  This argument does not apply to SF scattering because spin-$\bar{\sigma}$ electrons are not Pauli-blocked.  This explains the observed difference in SF versus NSF lineshapes as well as the insensitivity of NSF lineshapes to the core hole.  Because this effect is non-perturbative an exact analysis is indispensable for detecting it.  In the absence of a core hole potential the RIXS intensity can be calculated in the energy domain.  We obtain
\begin{align}
I \propto & \sum_{\alpha,\beta} \left|
\frac{ \sum_\vk \chi_{\rho \sigma} \braket{\alpha}{\vk+\vQ,\rho} \braket{\vk, \sigma}{\beta}}{\omega-\epsilon_\beta + i \Gamma} \right|^2 \nonumber \\ & \times n_f(\epsilon_\alpha)(1-n_f(\epsilon_\beta)) \delta(\epsilon_\alpha-\epsilon_\beta-\Delta \omega)
\label{SimpleDirect}
\end{align}
where $\ket{\vk,\vk+\vQ}$ are momentum eigenstates and $\ket{\alpha,\beta}$ are single-particle eigenstates of $H$ in the spin-orbital basis.

Finally, we note that the overall intensity of calculated and measured lineshapes changes very little from the optimally-doped to the overdoped material.  The experimental and theoretical lineshapes in Fig.~\ref{Lineshapes} were aligned by a \textit{single} factor for all momenta and both dopings.  If the RIXS signal came predominantly from a collective mode this would imply a spectral weight that varies little with doping.  In our model, however, this happens naturally because a slight change in chemical potential does not strongly affect the results.

{\it Summary and Outlook}.--  We derived a formalism to treat band structures, pairing, and core hole potentials in direct and indirect RIXS.  The lineshapes we calculated in a Fermi liquid-like model agreed well with experiments on cuprates over a wide range of doping for both spin-flip and non-spin-flip scattering, and we found a mechanism by which the core hole differentiates the two channels.  We concluded that dispersing peaks seen in RIXS experiments on cuprates may be attributable to band structure alone, rather than collective modes.  Thus the constant intensity of peaks in RIXS as doping increases does not imply a constant spectral weight of magnetic excitations, which has important implications for the mechanism of superconductivity in these materials~\cite{Scalapino2012}.

Our model of non-interacting quasiparticles is a priori well-supported by experimental evidence for the overdoped cuprates~\cite{Norman1998}.  The agreement of our model with measured data suggests that it remains valid to doping at least as low as $p=0.15$.  We expect that it would work as far as $p = 0.08$, where a Fermi surface is found in experiments~\cite{LeBoeuf2007a, Doiron-Leyraud2007a, Vignolle2011d}.  However, a non-interacting model becomes insufficient at some point in the underdoped regime.  The analysis presented in this paper can be extended to deeply-underdoped antiferromagnetic states via an RPA-like analysis, which is known to correctly reproduce spin wave excitations in the insulating state.  As the insensitivity of RIXS lineshapes to doping in the range we have considered persists to some extent to the undoped Mott antiferromagnet, a theory that bridges these two limits is very desirable.  

\begin{figure}
\includegraphics[width=0.9\linewidth]{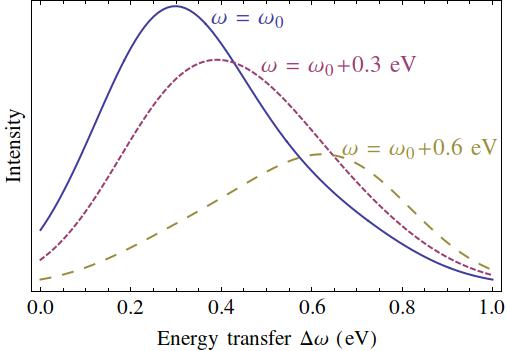} 
\caption{Calculated intensity vs. energy transfer for spin-flip RIXS of optimally-doped Bi-2212 (p=0.15) for antinodal momentum $\vQ = (\pi/a, 0)$ at incident energies 0, 0.2, and 0.4 eV above the absorption maximum.  The increase in $\Delta \omega$ with $\omega$ occurs when the RIXS final state belongs to the particle-hole continuum and does not occur if the final state is an excitation of a collective mode.} 
\label{incident}
\end{figure}

While we have shown that a model of non-interacting quasiparticles is in excellent agreement with RIXS experiments, we have not presented direct evidence to reject an interpretation in terms of collective modes.  Such evidence, however, could easily be obtained by measuring RIXS lineshapes for incident energy $\omega$ above the absorption maximum.  As shown in Fig.~\ref{incident}, as $\omega$ increases lineshapes move to larger $\Delta \omega$.  The RIXS signal due to inelastic scattering of a collective mode does not behave this way because $\Delta \omega$ cannot exceed the energy of the mode.  This brings up the important lesson that analyses of RIXS must calculate the RIXS signal itself, and not a proxy such as magnetic suceptibility.  A susceptibility $\chi(\omega, \vk)$ depends on a single frequency $\omega$, which corresponds to the energy of excitations.  Since the energy transfer in RIXS plays a similar role the correspondence $\Delta \omega$ (RIXS) $\rightarrow \omega$ (susceptibility) is often assumed.  However, this correspondence neglects the significant interplay of $\omega$ and $\Delta \omega$ in RIXS.  In RIXS the phase space for final states is modified by the intermediate state resonance.  For example, in Fig.~\ref{incident} the intermediate state photoelectron's energy increases with $\omega$, which tends to increase the energy of the final state particle-hole pair.

{\it Acknowledgements}.--  We acknowledge Peter Abbamonte for spurring our interest in RIXS, Mark Dean and Mathieu LeTacon for sharing data and helpful feedback, and Dmitry Abanin, Jeroen van den Brink, Marco Grioni, and Henrik Ronnow for discussions.  We acknowledge support from Harvard-MIT CUA, the ARO-MURI on Atomtronics, and the ARO MURI Quism program.
\appendix

\section{Appendix: Direct RIXS}
The averages $S_{mn}$ at finite temperatures have the form
\begin{align}
S^{mn}_{\rho \sigma \mu \nu} &  =  
\tr \left[ e^{i H \tau} d_p e^{-i H_n \tau} d^\dg_q e^{i H s} d_r  \ldots \right. \nonumber \\
\label{DirectTrace}
& \left. \ldots e^{i H_m t}  d^\dg_s e^{-i H(t+s)-\beta H}  \right]/\tr \left[ e^{-\beta H} \right], 
\end{align}
where $p,q,r,s$ are combined site and spin indices in a spin-Wannier basis, eg. $\ket{r}=\ket{m,\mu}$ in Eq. \ref{S}.  From the identity $\tr \, e^X = \det(1+e^x)$, the denominator of Eq.~(\ref{DirectTrace}) is $\det(1+e^{-\beta h})$.  We define $X_1 = \left( i (\tau - t - s) - \beta \right) H_0$, $X_2 = i H_m t$, $X_3 = i H_0 s$, $X_4 = - i H_n \tau$ and use the cyclicity property to express the numerator $N$ as $N=\tr \left[ d_p e^{X_4} d^\dg_q e^{X_3} d_r e^{X_2} d^\dg_s e^{X_1} \right]$.  Switching to an arbitrary basis of spin-orbitals, with implicit summation over Greek indices, gives
\begin{equation}
N=\braket{p}{\alpha} \braket{\beta}{q} \braket{r}{\gamma} \braket{\delta}{s} \tr \left[ d_\alpha e^{X_4} d^\dg_\beta e^{X_3} d_\gamma e^{X_2} d^\dg_\delta e^{X_1} \right].
\end{equation}
We move all $d/d^\dg$ to the left as follows.  Choose $\beta$ to be eigenstates of $X_4$ with eigenvalues $\omega_\beta$.  Then $\bra{\beta} \twodots e^{X_4} d^\dg_\beta = \bra{\beta} \twodots e^{\omega_\beta} d^\dg_\beta e^{X_4} = \bra{\beta} e^{X_4} \twodots  d^\dg_\beta e^{X_4}$.  After absorbing the c-number $e^{\omega_\beta}$ as $\bra{\beta} e^{X_4}$, the basis $\beta$ is again arbitrary.  The general pattern is to commute  $d^\dg_\delta$ with $e^X$ via $\bra{\delta} \rightarrow \bra{\delta} e^X$ and commute $d_\gamma$ with $e^X$ via $\ket{\gamma} \rightarrow e^{-X} \ket{\gamma}$.  Successive applications yield
\begin{align}
N =& \braket{p}{\alpha} \braketop{\beta}{e^{X_4}}{q} \braketop{r}{e^{-X_3}e^{-X_4}}{\gamma} \braketop{\delta}{e^{X_4} e^{X_3} e^{X_2}}{s} 
 \nonumber \\
& \times \tr \left[ d_\alpha d^\dg_\beta d_\gamma d^\dg_\delta e^Z \right],
\label{AllToTheLeft}
\end{align}
where $Z$ is a quadratic operator such that $e^Z = e^{X_4}  e^{X_3}  e^{X_2}  e^{X_1}$ whose existence is guaranteed by the Baker-Campbell-Haussdorff lemma.  If there were no $d/d^\dg$ insertions we would evaluate the trace in an eigenbasis $Z \ket{\phi} = \omega_\phi \ket{\phi}$, in which it factorizes as $\tr \, e^Z = \prod_\phi \sum_{n_\phi = 0, 1} e^{n_\phi \omega_\phi} = \prod_\phi (1 + e^{\omega_\phi}) = \det (1+e^z)$.  Now choose $\alpha,\beta,\gamma,\delta$ to be eigenstates of $Z$ with eigenvalues $\omega_\alpha$ etc.  The trace vanishes unless the $d/d^\dg$ match pairwise.  For example, if $\alpha = \beta \ne \gamma = \delta$ we have $\tr \, \left[(1-n_\alpha)(1-n_\gamma)e^Z \right]$.  This is identical to $\tr \, e^Z$ except the factor $(1+e^{\omega_{\alpha}})$ is replaced by $\sum_{n_{\alpha} = 0,1} (1-n_{\alpha})e^{n_\alpha \omega_\alpha}=1$, and similarly for $\gamma$ giving $ (1+e^{\omega_\alpha})^{-1} (1+e^{\omega_\gamma})^{-1} \det (1+e^z)$.  Including the other cases $\alpha = \delta \ne \beta = \gamma$ and $\alpha=\beta=\gamma=\delta$ we get
\begin{align}
\tr \, & \left[ d_\alpha d^\dg_\beta d_\gamma d^\dg_\delta e^Z \right]
= \det (1+e^Z) \left[ \delta_{\alpha\beta}\delta_{\gamma\delta}\delta_{\beta\gamma} (1+e^{\omega_\alpha})^{-1} \right.
 \nonumber \\
&+\delta_{\alpha\delta}\delta_{\beta\gamma}(1-\delta_{\alpha\beta}) (1+e^{\omega_\alpha})^{-1} e^{\omega_\beta} (1+e^{\omega_\beta})^{-1} \nonumber \\
&+\left. \delta_{\alpha\beta}\delta_{\gamma\delta}(1-\delta_{\alpha\gamma}) (1+e^{\omega_\alpha})^{-1} (1+e^{\omega_\gamma})^{-1} \right] \nonumber \\
&= \det (1+e^Z) \left[ 
\delta_{\alpha\delta}\delta_{\beta\gamma}  (1+e^{\omega_\alpha})^{-1} e^{\omega_\beta} (1+e^{\omega_\beta})^{-1} \right. \nonumber \\
&+\left. \delta_{\alpha\beta}\delta_{\gamma\delta} (1+e^{\omega_\alpha})^{-1} (1+e^{\omega_\gamma})^{-1} \right].
\label{TraceResult}
\end{align}
Next, we absorb the c-numbers in Eq.~\ref{TraceResult} as operators via $\ket{\alpha} \twodots (1+e^{\omega_\alpha})^{-1} = (1+e^Z)^{-1}\ket{\alpha}$ and apply the Kronecker $\delta$'s via eg. $\braketop{p}{\twodots}{\alpha}\braketop{\beta}{\twodots}{q} \delta_{\alpha\beta}=\braketop{p}{\twodots}{q}$, obtaining
\begin{align}
S_{pqrs}=&\frac{\det (1+e^Z)}{\det (1+e^{-\beta H_0})} \times \left[
\braketop{p}{(1+e^{x_5})^{-1}e^{x_4}}{q} \right. \nonumber \\
& \times \braketop{r}{e^{-x_3}e^{-x_4}(1+e^{x_5})^{-1}e^{x_4}e^{x_3}e^{x_2}}{s}
             \nonumber \\          
 &+ \braketop{p}{(1+e^{x_5})^{-1}e^{x_4}e^{x_3}e^{x_2}}{s} \nonumber \\
  \label{Ugly} 
  &   \times \left. \braketop{r}{e^{-x_3}e^{-x_4}e^{x_6} (1+e^{x_5})^{-1} e^{x_4}}{q}
 \right] 
\end{align}

Eq.~(\ref{Ugly}) becomes more physical upon introducing $U_{mn} = e^{x_4}e^{x_3}e^{x_2} = e^{-i h_n \tau}e^{i h s} e^{i h_m t}$ and $U_0 = e^{i(\tau-t-s)h}$, which are Keldysh propagators with and without core holes.  Additionally, we rewrite $e^{-\beta h_0} = N/(1-N)$ in terms of occupation operators.  Then we have $(1+e^{x_5})^{-1} = (1-N)(1-N + U_{mn} U_0 N)^{-1}=(1-N)F^{-1}$, where $F = 1-N + U_{mn} U_0 N$ gives the overlap of core-hole and core-hole-less propagation of initially occupied states.  The ratio of determinants comes out to $\det (1+e^z)/\det(1+e^{-\beta h}) = \det (F)$ and we obtain, after restoring $p = n, \rho; q = n, \sigma; r = m, \mu; s = m, \nu$,
 \begin{align}
 &S^{mn}_{\rho \sigma \mu \nu} = \det(F) \left[
 \braketop{n\rho}{(1-N)F^{-1} e^{-i h_n \tau}}{n\sigma} \right. \nonumber \\
 & \times \braketop{m\mu}{e^{-i h s} e^{i h_n \tau} (1-N) F^{-1} U_{mn}}{m\nu}   \nonumber \\
 &+ \braketop{n\rho}{(1-N)F^{-1} U_{mn}}{m\nu} \nonumber \\
&\times \left. \braketop{m\mu}{e^{i h_m t} U_0 N F^{-1} e^{-i h_n \tau}}{n\sigma} \right].
\label{SpinfulResult}
 \end{align}
%So far this all pertains to the full Hilbert space of spin-orbitals.  If $H$ contains no spin-flip terms $S^{mn}_{\rho \sigma \mu \nu}$ simplifies.  The Fermi sea term $\det F$ factorizes as $\det F_\up \det F_\dn$.  SF processes have $\sigma = \bar{\rho}$ and $\mu=\bar{\nu}$, whence the term $\braketop{n\rho}{\ldots}{n\sigma}$ vanishes.  Thus all matrices need pertain only to spinless Hilbert space, not the spin-doubled Hilbert space.
%\begin{align}
%S^{\rm cons}_{mn} =  \det & (F_\up F_\dn) \sum_\sigma  \braketop{n}{(1-N_\sigbar)F_\sigbar^{-1} U^\sigbar_{mn}}{m} \nonumber \\
 %& \times \braketop{m}{e^{i H^\sigma_m t} U^\sigma_0 N_\sigma F_\sigma^{-1} e^{-i H^\sigma_n \tau}}{n}
%\end{align}
%for the spin-flip intensity and
%\begin{align}
%S^{\rm flip}_{mn} =  \det & (F_\up F_\dn) \left\{ \sum_\sigma  \braketop{n}{(1-N_\sigma)F_\sigma^{-1} U^\sigma_{mn}}{m} \right. \nonumber \\
%& \times \braketop{m}{e^{i H^\sigma_m t} U^\sigma_0 N_\sigma F_\sigma^{-1} e^{-i H^\sigma_n \tau}}{n}  \nonumber \\
%&+ \sum_{\sigma,\sigprime}  \braketop{n}{(1-N_\sigprime)F_\sigprime^{-1} e^{-i H^\sigprime_n \tau}}{n} \nonumber \\
%&\times \left.  \braketop{m}{e^{i H^\sigma_m t} U^\sigma_0 N_\sigma F_\sigma^{-1} e^{-i H^\sigma_n \tau}}{m} \right\}
%\end{align}
%for the spin-conserving intensity.  If $H_0$ is spin-independent the sums $\sum_\sigma$ and $\sum_{\sigma,\sigprime}$ reduce to factors of 2 and 4.

\bibliography{library}

\end{document}

\title{Supplemental Material: Explicit Derivation of Spin-Orbit Effects}
\maketitle

Here we provide an explicit derivation of the separation of RIXS scattered intensity into spin-flip and spin-conserving channels due to spin-orbit coupling a the $2p$ core hole.  For earlier discussions see Refs.~\cite{Ament2009, Ament2011, Haverkort2010}.  We calculate the polarization dependence of the two channels and verify that experiments with grazing-exit geometry afford a very clean separation.  In doing so, we justify the form of dipole transition operators $T/T^\dg$ discussed in the text.
 
\section{Dipole Matrix Elements}
In the electric dipole approximation, matrix elements for light of polarization $\hat{\epsilon}$ to cause an electronic transition from state $\ket{\psi_i, \sigma_i}$ to $\ket{\psi_f, \sigma_f}$ is
\begin{equation}
\hat{\epsilon} \cdot \braketop{\psi_f}{\vr}{\psi_i} \delta_{\sigma_i, \sigma_f}.
\end{equation}
At the cuprate $L3$ edge we are interested in $\ket{\psi_i} = \ket{2p; m = -1, 0, 1}$ and $\ket{\psi_f} = \ket{3d_{x^2-y^2}}$.  
Dipole matrix elements are most easily evaluated in the $\{ p_x, p_y, p_z \}$ basis, where by symmetry we can easily see that the only non-zero matrix elements are
\begin{equation}
\braketop{3d_{x^2-y^2}}{x}{2p_x} = - \braketop{3d_{x^2-y^2}}{y}{2p_y}.
\end{equation}
We switch from the $\ket{2p_{x, y, z}}$ basis to the $\ket{m_\ell = -1, 0, 1}$ basis via
\begin{equation}
\ket{m_\ell = 0} = \ket{2p_z}, \, \ket{m_\ell = \pm 1} = \frac{1}{\sqrt{2}} \left( \ket{2p_x} \pm i \ket{2p_y} \right)
\end{equation}
to obtain the dipole matrix elements (up to an overall multiplicative constant)
\begin{equation}
\hat{\eta}_m \equiv \braketop{3d_{x^2-y^2}}{\vr}{2p_m} = \left\{
\begin{array}{cc}
0 & (m=0) \\
\frac{1}{\sqrt{2}} \left( \hat{x} \mp i \hat{y} \right) & (m = \pm 1)
\end{array}
\right.
\end{equation}

\section{Spin-Orbit Basis}
To include the spin-orbit effect, we must be able to translate between the basis $\ket{m_\ell = -1, 0, 1; \sigma = \up, \dn}$, which is most convenient for deriving dipole matrix elements, to the basis $\ket{j, m_j}$ of energy eigenstates.  From a Clebsch-Gordan table we find
\begin{align}
\ket{3/2, 3/2} &= \ket{1, \up} \\
\ket{3/2, 1/2} &= \sqrt{1/3} \ket{1, \dn} + \sqrt{2/3} \ket{0, \up} \\
\ket{3/2, -1/2} &= \sqrt{2/3} \ket{0, \dn} + \sqrt{1/3} \ket{-1, \up} \\
\ket{3/2, -3/2} &= \ket{-1, \dn}.
\end{align}
We invert this and drop terms from the $\ket{j=1/2}$ subspace, which are off resonance by $\sim$ 20 eV at the $L3$ edge.  This omission of $\ket{j=1/2}$ terms breaks spin rotational symmetry and allows for spin-flip RIXS.  We find for the $\ket{m_\ell = \pm 1}$ states (as seen above, $\ket{m_\ell = 0}$ has vanishing matrix elements)
\begin{align}
\ket{1, \up} &= \ket{3/2, 3/2} \\
\ket{1, \dn} &= \sqrt{1/3} \ket{3/2, 1/2} \\
\ket{-1, \up} &= \sqrt{1/3} \ket{3/2, -1/2} \\
\ket{-1, \dn} &= \ket{3/2, -3/2}
\end{align}

\section{Transition Operator}
The Kramers-Heisenberg amplitude for initial state $\ket{i}$ and final state $\ket{f}$ is
\begin{align}
A_{i \rightarrow f} &= \sum_n \frac{ \braketop{f}{T_f}{n} \braketop{n}{T^\dg_i}{n} } { \omega - E_n + i \Gamma } \\
&= \braketop{f}{ T_f G T^\dg_i } {i},
\end{align}
where $T^\dg_i$ and $T_f$ are absorption and emission dipole transition operators and $G \equiv (\omega - H + i \Gamma)^{-1}$ is the intermediate state Green function.  The transition operators are obtained by adding dipole absorption/emission events for all possible $\sigma$, $m_\ell$, and core hole sites $\vR$:
\begin{align}
T^\dg_i =& \sum_{\vR, m, \sigma} e^{-i \vq_i \cdot \vR} \hat{\epsilon}_i \cdot \hat{\eta}_m d^\dg_{\vR, \sigma} p_{\vR, \sigma, m} \\
T_f =& \sum_{\vR, m, \sigma} e^{i \vq_f \cdot \vR} \hat{\epsilon}^*_f \cdot \hat{\eta}^*_m  p^\dg_{\vR, \sigma, m} d_{\vR, \sigma},
\end{align}
where $\hat{\epsilon}_{i(f)}$ and $\vq_{i(f)}$ are incident (final) polarizations and photon momenta.  Writing this in terms of spin-orbit eigenstates and dropping off-resonance $\ket{j=1/2}$ terms as in the previous section gives
\begin{align}
T^\dg_i =& \sum_\vR e^{-i \vq_i \cdot \vR} \hat{\epsilon}_i \left[
\hat{\eta}_1 \left( d^\dg_{\vR, \up} p_{\vR,m_j=3/2} +\sqrt{1/3} d^\dg_{\vR,\dn} p_{\vR, m_j = 1/2} \right) \right. \nonumber \\
&  + \left. \hat{\eta}_{-1} \left( d^\dg_{\vR, \dn} p_{\vR,m_j=-3/2} +\sqrt{1/3} d^\dg_{\vR,\up} p_{\vR, m_j = -1/2} \right) \right] \\
T_f =& \sum_\vR e^{i \vq_f \cdot \vR} \hat{\epsilon}^*_f \left[
\hat{\eta}^*_1 \left( d^\dg_{\vR, \up} p_{\vR,m_j=3/2} +\sqrt{1/3} d^\dg_{\vR,\dn} p_{\vR, m_j = 1/2} \right) \right. \nonumber \\
& + \left. \hat{\eta}^*_{-1} \left( d^\dg_{\vR, \dn} p_{\vR,m_j=-3/2} +\sqrt{1/3} d^\dg_{\vR,\up} p_{\vR, m_j = -1/2} \right) \right].
\end{align}
Finally, since the spin orbit states labelled by $\{\vR, j, m_j \}$ are eigenstates of $H$, each term in $T^\dg_i$ must be matched with the corresponding term in $T_f$.  Thus we obtain
\begin{equation}
A_{i \rightarrow f} = \sum_\vR e^{i \Delta \vq \cdot \vR} \chi_{\alpha, \beta}
\braketop{f}{d_{\vR,\alpha} G d^\dg_{\vR, \beta}}{i},
\end{equation}
where
\begin{align}
\chi_{\up, \up} = \chi_{\dn, \dn}^* =& (\hat{\epsilon_f} \cdot \hat{\eta}_1)^*(\hat{\epsilon_i} \cdot \hat{\eta}_1)
+ (1/3) (\hat{\epsilon_f} \cdot \hat{\eta}_{-1})^*(\hat{\epsilon_i} \cdot \hat{\eta}_{-1}) \\
\chi_{\up, \dn} = \chi_{\dn, \up} =& 0.
\end{align}
That $\chi_{\up,\up} = \chi_{\dn,\dn}^*$ follows from $\hat{\eta}_1 = \hat{\eta}_{-1}^*$.

\section{Spin-Flip and Non-Spin-Flip Channels}
To separate the spin and charge channels, we decompose $\chi$ into symmetric and antisymmetric parts: $\chi = \chi_S + \chi_A$, where 
\begin{align}
\chi_S =& (\chi + \chi^*)/2 \\
\chi_A =& (\chi - \chi^*)/2.
\end{align}
Then, in the basis of $\hat{S}_z$ eigenstates $\ket{\up, \dn}$, where $\hat{z}$ is  perpendicular to the copper-oxide plane,  we have
\begin{align}
\left( \chi_S \right)_{\up,\up} = \left( \chi_S \right)_{\dn,\dn} =& (2/3)\left( \hat{\epsilon}_f^* \cdot \hat{\epsilon}_i - \epsilon_{f,z}^* \epsilon_{i,z} \right) &  \equiv \chi_S \\
\left( \chi_A \right)_{\up,\up} = -\left( \chi_A \right)_{\dn,\dn} =& -(i/3) \hat{\epsilon}_f^* \cdot (\hat{z} \times \hat{\epsilon}_i ) & \equiv \chi_A
\end{align}
With respect to the basis of eigenstates of $\hat{S}_x$, obtained via the transformations $\ket{\up} \rightarrow (1/\sqrt{2}) ( \ket{\up} + \ket{\dn} )$ and $\ket{\dn} \rightarrow (1/\sqrt{2}) ( \ket{\up} - \ket{\dn} )$, the symmetric amplitude transforms trivially:
\begin{equation}
\chi_S d_\up G d^\dg_\up + \chi_S  d_\dn G d^\dg_\dn \rightarrow \chi_S \left( d_\up G d^\dg_\up +  d_\dn G d^\dg_\dn \right) .
\end{equation}
An isotropic contribution can't represent a spin flip in any basis.  The antisymmetric contribution, however, transforms as
\begin{equation}
\chi_A d_\up G d^\dg_\up - \chi_A  d_\dn G d^\dg_\dn \rightarrow \chi_A \left( d_\up G d^\dg_\dn +  d_\dn G d^\dg_\up \right) ,
\end{equation}
a pure spin flip.  Now we need to compare the intensity prefactors $|\chi_{S(A)}|^2$.

\section{Evaluation for Specific Geometries}
For experimental geometry where incident and scattered radiation make angles $\phi_1$ and $\phi_2$ with $\hat{z}$ polarization vectors are $\hat{\epsilon}_i = \hat{\epsilon}_f = \hat{y}$ for $\sigma$ polarization (WLOG) and $\hat{\epsilon}_i = \cos \phi_1 \hat{x} + \sin \phi_1 \hat{z}$, $\hat{\epsilon}_f = \cos \phi_2 \hat{x} + \sin \phi_2 \hat{z}$ for $\pi$ polarization.  Then we can calculate $\chi_S$ and $\chi_A$ for all pairs of incident and emitted polarizations:
\begin{equation}
\begin{array}{cccc}
\hat{\epsilon}_i & \hat{\epsilon}_f & \chi_S & \chi_A \\
\sigma & \sigma & 2/3 & 0 \\
\sigma & \pi & 0 & (i/3) \cos \phi_2 \\
\pi & \pi & (2/3) \cos \phi_1 \cos \phi_2 & 0 \\
\pi & \sigma & 0 & -(i/3) \cos \phi_1
\end{array}
\end{equation}
For unpolarized scattered radiation we average intensity over scattered polarizations:
\begin{equation}
\begin{array}{ccc}
\hat{\epsilon}_i & \ave{|\chi_S|^2} & \ave{|\chi_A|^2} \\
\sigma & 2/9 & (1/18) \cos^2 \phi_2 \\
\pi & (2/9) \cos^2 \phi_1 \cos^2 \phi_2  & (1/18) \cos^2 \phi_1
\end{array}
\end{equation}
The prefactors $|\chi_S(\hat{\epsilon}_i, \hat{\epsilon}_f)|^2$ and $|\chi_S(\hat{\epsilon}_i, \hat{\epsilon}_f)|^2$ depend only on the polarizations and as such do not affect the lineshape of intensity vs. momentum transfer and energy transfer.  This means that we can write
\begin{align}
I^{\rm total}_\sigma =& |\chi_{S,\sigma}|^2 I^{\rm NSF} + |\chi_{A,\sigma}|^2 I^{\rm SF} \\
I^{\rm total}_\pi =& |\chi_{S,\pi}|^2 I^{\rm NSF} + |\chi_{A,\pi}|^2 I^{\rm SF}.
\end{align}
Even without resolving the polarization of scattered radiation, the charge and spin channels can, in principle, be separated by solving a $2 \times 2$ linear equation.  Ideally, however, one could simply choose geometries where $\ave{|\chi_S|^2}$ is much larger than $\ave{|\chi_A|^2}$ and vice versa.

We consider the experiments of Refs.~\cite{Dean2013, LeTacon2013}, with grazing exit geometry.  For concreteness we use values from Ref.~\cite{Dean2013}: $\phi_i = -25.6^\circ$ and $\phi_f = 76.6^\circ$ (the minus sign denotes that incident and scattered radiation are on the same side of the normal in the scattering plane).  For this geometry, averaging over final states we get
\begin{equation}
\begin{array}{ccc}
\hat{\epsilon}_i & \ave{|\chi_S|^2} & \ave{|\chi_A|^2} \\
\sigma & 0.22 & 0.003\\
\pi & 0.010  & 0.045
\end{array}.
\end{equation}

Thus the two polarizations of incident radiation in an experiment with grazing exit geometry offer a fairly clean separation between spin-flip and spin-conserving cross sections.

\bibliography{library}